\documentclass[aps,prd,twocolumn,groupedaddress,showpacs,nofootinbib,amssymb]{revtex4}
\usepackage[dvips]{graphicx}
\usepackage{amssymb}
\usepackage{amsmath}
\usepackage{graphicx,,color}
\usepackage{amsfonts}
\usepackage{bm}
\usepackage{cancel}
\usepackage{comment}

\newcommand\be{\begin{equation}}
\newcommand\ee{\end{equation}}

\allowdisplaybreaks[4]

\begin{document}

\tolerance=5000

\title{Generalizing the Constant-roll Condition in Scalar Inflation}
\author{V.K.~Oikonomou,$^{1,2,3}$\,\thanks{v.k.oikonomou1979@gmail.com}}

\affiliation{$^{1)}$ Department of Physics, Aristotle University
of Thessaloniki, Thessaloniki 54124,
Greece\\
$^{2)}$ Laboratory for Theoretical Cosmology, Tomsk State
University of Control Systems
and Radioelectronics, 634050 Tomsk, Russia (TUSUR)\\
$^{3)}$ Tomsk State Pedagogical University, 634061 Tomsk, Russia\\
}

\tolerance=5000

\begin{abstract}
In this work we generalize the constant-roll condition for
minimally coupled canonical scalar field inflation. Particularly,
we shall assume that the scalar field satisfies the condition
$\ddot{\phi}=\alpha (\phi) V'(\phi)$, and we derive the field
equations under this assumption. We call the framework extended
constant-roll framework. Accordingly we calculate the inflationary
indices and the corresponding observational indices of inflation.
In order to demonstrate the inflationary viability, we choose
three potentials that are problematic in the context of slow-roll
dynamics, namely chaotic, linear power-law and exponential
inflation, and by choosing a simple power-law form for the smooth
function $\alpha (\phi)$, we show that in the extended
constant-roll framework, the models are compatible with the latest
2018 Planck constraints on inflation. We also justify
appropriately why we called this new framework extended
constant-roll framework, and we show that the condition
$\ddot{\phi}=\alpha (\phi) V'(\phi)$ is equivalent to the
condition $\ddot{\phi}=\beta (\phi) H \dot{\phi}$, with the latter
condition being a simple generalization of the constant-roll
condition. Finally, we examine an interesting physical situation,
in which a general extended constant-roll scalar field model is
required to satisfy the cosmological tracker condition used in
quintessence models. In contrast to the slow-roll and ordinary
constant-roll cases, in which case the tracker condition is not
compatible with neither the slow-roll or the ordinary
constant-roll conditions, the extended constant-roll condition can
be compatible with the tracker condition. This feature leads to a
new inflationary phenomenological framework,  the essential
features of which we develop in brief. The main feature of the new
theoretical framework is that the function $\alpha (\phi)$ and the
potential $V(\phi)$ are no longer free to choose, but these are
directly functionally related.
\end{abstract}


\maketitle

\section{Introduction}

The most common description of the inflationary era is
materialized by a canonical minimally coupled scalar field
\cite{Guth:1980zm,Starobinsky:1982ee,Linde:1983gd,Albrecht:1982wi},
called inflaton. To date, it has not be verified that inflation
ever took place primordially, nevertheless inflation solves
smoothly many theoretical inconsistencies of the standard
cosmological model, thus it remains the most appealing candidate
for the description of the primordial post-Planck era, with the
most promising alternative class of theories being modified
gravity theories
\cite{Nojiri:2017ncd,Nojiri:2010wj,Nojiri:2006ri,Capozziello:2011et,Capozziello:2010zz,delaCruzDombriz:2012xy,Olmo:2011uz}.
Nowadays, the latest Planck data \cite{Akrami:2018odb} impose
stringent constraints on the inflationary parameters, and it seems
that the power-spectrum of the primordial curvature perturbations
is nearly scale invariant. The standard scalar field description
of inflation assumes that the scalar field slowly rolls down its
potential, however during the last decades, several alternative
descriptions like the constant-roll case
\cite{Inoue:2001zt,Tsamis:2003px,Kinney:2005vj,Tzirakis:2007bf,
Namjoo:2012aa,Martin:2012pe,Motohashi:2014ppa,Cai:2016ngx,
Motohashi:2017aob,Hirano:2016gmv,Anguelova:2015dgt,Cook:2015hma,
Kumar:2015mfa,Odintsov:2017yud,Odintsov:2017qpp,Lin:2015fqa,Gao:2017uja,Nojiri:2017qvx,Oikonomou:2017bjx,Odintsov:2017hbk,Oikonomou:2017xik,Cicciarella:2017nls,Awad:2017ign,Anguelova:2017djf,Ito:2017bnn,Karam:2017rpw,Yi:2017mxs,Mohammadi:2018oku,Gao:2018tdb,Mohammadi:2018wfk,Morse:2018kda,Cruces:2018cvq,GalvezGhersi:2018haa,Boisseau:2018rgy,Gao:2019sbz,Lin:2019fcz}
have been studied in the literature. In this paper we aim to
generalize the constant-roll condition in such a way so that
$\ddot{\phi}=\alpha (\phi)V'(\phi)$, where $\alpha (\phi)$ is some
smooth function of the scalar field, and we call this framework
the extended constant-roll framework. We develop the extended
constant-roll framework and derive expressions for the
inflationary indices and for the observational indices of
inflation, focusing on the spectral index of the scalar
perturbations and on the tensor-to-scalar ratio. In order to
examine the viability of the resulting framework, we choose three
popular inflationary models, which in the slow-roll inflation are
problematic, namely the chaotic inflation, the linear power-law
and the exponential inflation models. As we show, for a simple
power-law choice of the smooth function $\alpha (\phi)$, the three
models are in good agreement with the Planck data. Also we explain
why we named the condition $\ddot{\phi}=\alpha (\phi)V'(\phi)$ as
extended constant-roll condition, since this condition is
equivalent to $\ddot{\phi}=\beta (\phi) H \dot{\phi}$, by
employing a simple transformation. Finally, we demonstrate that
the imposition of the tracker condition is compatible with the
extended constant-roll scenario, a feature of the theoretical
framework we developed, which can play an important role for the
unified description of inflation with the dark energy era, by
using a single scalar field.

For the rest of the study, we shall use a flat
Friedman-Robertson-Walker (FRW) spacetime, with its line element
being,
\begin{equation}
\label{metricfrw}
ds^2 = - dt^2 + a(t)^2 \sum_{i=1,2,3} \left(dx^i\right)^2\, ,
\end{equation}
where $a(t)$ is as usual the scale factor.

\section{Extended Constant-roll: Formalism and Applications}

We shall consider minimally coupled scalar field cosmology, with
the action being,
\begin{equation}\label{action}
\mathcal{S}=\int
d^4x\left(\frac{R}{2\kappa^2}-\frac{1}{2}\partial_{\mu}\phi\partial^{\mu}\phi-V(\phi)
\right)\, ,
\end{equation}
where $\kappa^2=\frac{1}{M_p^2}$ where $M_p$ is the reduced Planck
mass. The field equations corresponding to the above action are,
\begin{equation}
\label{motion1a} \centering 3H^2=\kappa^2\left(
\frac{1}{2}\dot\phi^2+V\right)\, ,
\end{equation}
\begin{equation}
\label{motion2a} \centering -2\dot H= \kappa^2\dot\phi^2\, ,
\end{equation}
\begin{equation}
\label{motion3a} \centering \ddot{\phi}+3H\dot{\phi}+V'=0\, .
\end{equation}
The slow-roll indices for a minimally coupled scalar theory are
\cite{Hwang:2005hb},
\begin{align}\label{slowrollindicesdef}
& \epsilon_1=-\frac{\dot H}{H^2}, \\ \notag &
\epsilon_2=\frac{\ddot\phi}{H\dot\phi},
\end{align}
and the observational indices for the inflationary theory, namely
the spectral index of scalar perturbations $n_s$ and the
tensor-to-scalar ratio are \cite{Nojiri:2017ncd,Hwang:2005hb},
\begin{equation}\label{spectralindexdef}
n_s=1-4 \epsilon_1-2\epsilon_2\, ,
\end{equation}
\begin{equation}\label{tensordef}
r=\frac{8\kappa^2Q_s}{f_R}\, ,
\end{equation}
where $Q_s=\frac{\dot{\phi}^2}{H^2}$. In view of the Raychaudhuri
equation (\ref{motion2a}) we get
$Q_s=-\frac{2\dot{H}}{\kappa^2H^2}$, therefore, the
tensor-to-scalar ratio reads,
\begin{equation}\label{final}
r=16 \epsilon_1\, .
\end{equation}
The extended constant-roll condition we shall consider in this
work is the following,
\begin{equation}\label{extendedconstantrollcondition1}
\ddot{\phi}=\alpha (\phi) V'(\phi)\, ,
\end{equation}
where $\alpha (\phi)$ is a smooth dimensionless function of the
scalar field $\phi$. In view of the extended constant-roll
condition (\ref{extendedconstantrollcondition1}), the equation of
motion for the scalar field (\ref{motion3a}) yields,
\begin{equation}\label{modifiedscalareqn}
\dot{\phi}=-\frac{V'(\phi)\left( 1+\alpha (\phi)\right)}{3 H}\, .
\end{equation}
Now assuming the standard slow-roll condition for the scalar
field, namely,
\begin{equation}\label{slowrollconditionscalarfield}
\frac{\dot{\phi}^2}{2}\ll V(\phi)\, ,
\end{equation}
the Friedmann equation yields as usual,
\begin{equation}\label{friedmannupdate}
\frac{3H^2}{\kappa^2}\simeq V(\phi)\, ,
\end{equation}
and upon combining Eqs. (\ref{modifiedscalareqn}) and
(\ref{friedmannupdate}), the slow-roll indices
(\ref{slowrollindicesdef}) take the following compact forms,
\begin{align}\label{slowrollindicesdef}
\epsilon_1=\frac{1}{2\kappa^2}\left(\frac{V'}{V}\right)^2\left(1+\alpha
(\phi) \right)^2, \,\,\, \epsilon_2=-\frac{3 \alpha
(\phi)}{1+\alpha (\phi)}\, .
\end{align}
The effect of the extended constant-roll condition is apparent on
the slow-roll indices, since $\epsilon_2$ becomes independent of
the potential, and the first slow-roll index depends explicitly on
$\alpha (\phi)$. Obviously, the standard constant-roll condition
can be obtained by considering $\alpha (\phi)=\mathrm{const}$, but
we shall extensively discuss this issue in a later section because
there exists a symmetry which can directly connect the extended
constant-roll condition (\ref{extendedconstantrollcondition1})
with the standard constant-roll condition. Let us proceed in
finding the expression for the $e$-foldings number for the case of
the extended constant-roll scenario, and we easily find by using
Eq. (\ref{modifiedscalareqn}) that,
\begin{equation}\label{efoldingsextendedconstantroll}
N=\int_{\phi_i}^{\phi_f}\frac{H}{\dot{\phi}}d\phi=\kappa^2\int_{\phi_f}^{\phi_i}\frac{V}{V'}\frac{1}{\left(
1+\alpha (\phi)\right )}d\phi\, ,
\end{equation}
where $\phi_i$ is the value of the scalar field at the first
horizon crossing primordially during the first stages of
inflation, and $\phi_f$ is the value of the scalar field at the
end of the inflationary era. Having Eqs. (\ref{spectralindexdef}),
Eqs. (\ref{final}), Eqs. (\ref{slowrollindicesdef}), and Eqs.
(\ref{efoldingsextendedconstantroll}), we can directly test
several cosmological scalar field models and confront the
inflationary phenomenology they produce with the 2018 Planck data
\cite{Akrami:2018odb}. In the next subsections we shall consider
three scalar field models which without the extended constant-roll
condition are not compatible with the 2018 Planck data, and thus
phenomenologically excluded. As we shall demonstrate, in the
context of the extended constant-roll scenario, these models
become compatible with the Planck observational constraints.

Before we close, let us in brief quote the formulas that yield the
predicted non-Gaussianities, just for the sake of completeness.
For the models studied in the next section, the predicted
non-Gaussianities are not reportable because these are quite
small. The formula that yields the equilateral parameter
$f_{NL}^{equil}$ can be found in Ref. \cite{DeFelice:2011zh}, and
it is using their notation,
\begin{equation}\label{equilpar}
f_{NL}^{equil}=\frac{11}{56}\epsilon_1+\frac{5}{12}\eta\, ,
\end{equation}
where $\eta=\frac{\dot{\epsilon}_1}{H\epsilon_1}$. In our
notation, the relation that connects $\eta$ and the slow-roll
indices $\epsilon_1$ and $\epsilon_2$ can easily be found by using
the equations of motion, and this is,
\begin{equation}\label{etarelation}
\epsilon_2=\frac{\ddot{H}}{2
H\dot{H}}=-\epsilon_1+\frac{\dot{\epsilon}_1}{2\epsilon_1 H}\, ,
\end{equation}
or equivalently,
\begin{equation}\label{etafinal}
\eta=2\left(\epsilon_1+\epsilon_2 \right)\, .
\end{equation}
Hence, the equilateral parameter $f_{NL}^{eqil}$ can be written in
terms of the slow-roll indices $\epsilon_1$ and $\epsilon_2$ as
follows,
\begin{equation}\label{fnlequilateralfinal}
f_{NL}^{equil}=\frac{55}{36}\epsilon_1+\frac{10}{12}\left(\epsilon_1+\epsilon_2\right)\,
.
\end{equation}

\subsection{Chaotic, Linear and Exponential Inflation Phenomenology with the Extended Constant-roll Condition}

We shall apply the extended constant-roll formalism we developed
previously in the case of chaotic inflation scenario
\cite{Linde:1983gd} which was quite popular in the 80's and 90's
but it is not viable with respect to the Planck data. As we now
show, by choosing a simple power law form for the dimensionless
function $\alpha (\phi)$, the chaotic inflation model can become
compatible with observations.
\begin{figure}[h!]
\centering
\includegraphics[width=22pc]{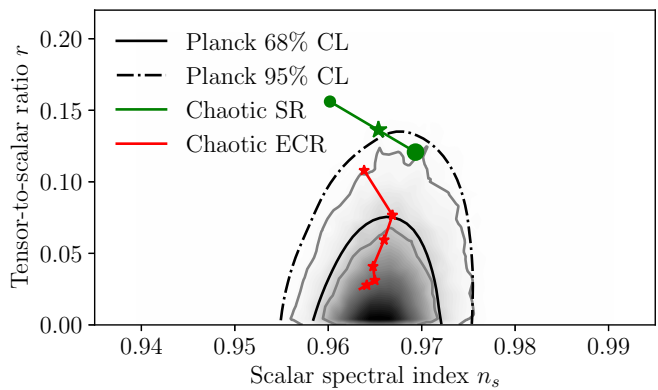}
\caption{The extended constant-roll chaotic inflation scenario
(red lines and points Extended Constant-roll (ECR)) versus the
ordinary slow-roll chaotic inflation scenario (green lines and
points, Slow-roll (SR)), in view of the Planck 2018 data.}
\label{plot1}
\end{figure}
The chaotic inflation scenario has the following scalar potential
\cite{Linde:1983gd} $V(\phi)=\frac{V_0}{\kappa^4} \left(\kappa
\phi \right)^2$, where $V_0$ is the dimensionless potential
multiplication factor. The upper value of $\frac{V_0}{\kappa^4}$
is determined by the latest Planck data, but it is irrelevant for
our analysis, so we do not discuss this issue further. The
dimensionless function $\alpha (\phi)$ entering in the extended
constant-roll condition (\ref{extendedconstantrollcondition1})
shall be chosen as $\alpha (\phi)=\delta \left(\kappa \phi
\right)^m-1$ The slow-roll indices can easily be calculated for
the above choices and these are $\epsilon_1=2 \delta ^2 \kappa ^{2
m-2} \phi ^{2 m-2}$, $\epsilon_2=\frac{3 \kappa ^{-m} \phi
^{-m}}{\delta }-3$. We can find the value of the scalar field at
the end of inflation $\phi_f$, by solving the equation
$\epsilon_1(\phi_f)=\mathcal{O}(1)$, and accordingly, by
performing the integration (\ref{efoldingsextendedconstantroll}),
we obtain the value of the scalar field at first horizon crossing
$\phi_i$, so these two are $\phi_f=\frac{2^{\frac{1}{2-2 m}}
\delta ^{\frac{1}{1-m}}}{\kappa }$, $\phi_i=2^{\frac{1}{2-m}}
\delta ^{\frac{1}{2-m}} \kappa ^{-1} \left(2^{\frac{m}{2-2 m}}
\delta ^{\frac{1}{1-m}}+(m-2) N\right)$. The spectral index of the
scalar curvature perturbations $n_s$ and the tensor-to-scalar
ratio $r$ acquire the following simple forms in terms of the
scalar field, $n_s= -8 \delta ^2 \kappa ^{2 m-2} \phi ^{2
m-2}-\frac{6 \kappa ^{-m} \phi ^{-m}}{\delta }+7$, and $r= 32
\delta ^2 \kappa ^{2 m-2} \phi ^{2 m-2}$. By evaluating the
slow-roll indices at first horizon crossing, thus for the scalar
field taking the value $\phi_i$ quoted above, we can easily
confront this model with the Planck 2018 observational data. The
results of our analysis are presented in Fig. \ref{plot1}. The
free parameters values are in the ranges $\delta=10^{-5}-10^{-19}$
and $m=4-12$, for $N=60$ $e$-foldings, and the extended
constant-roll chaotic inflation model corresponds to the red lines
and points in Fig. \ref{plot1}. We have also included in Fig.
\ref{plot1} the ordinary chaotic inflation model with green lines
and points. Obviously, the extended constant-roll chaotic
inflation model is well fitted with the Planck 2018 data, in
contrast to the ordinary chaotic inflation scenario.

\begin{figure}[h!]
\centering
\includegraphics[width=22pc]{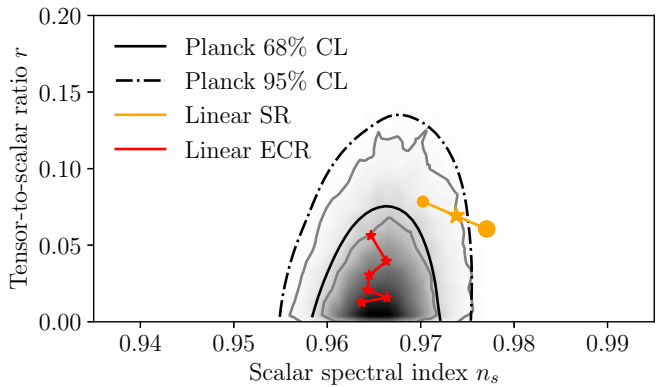}
\caption{The extended constant-roll linear power-law inflation
scenario (red lines and points  Extended Constant-roll (ECR))
versus the ordinary slow-roll linear power-law inflation scenario
(yellow lines and points, Slow-roll (SR)), in view of the Planck
2018 data.} \label{plot2}
\end{figure}

\begin{figure}[h!]
\centering
\includegraphics[width=18pc]{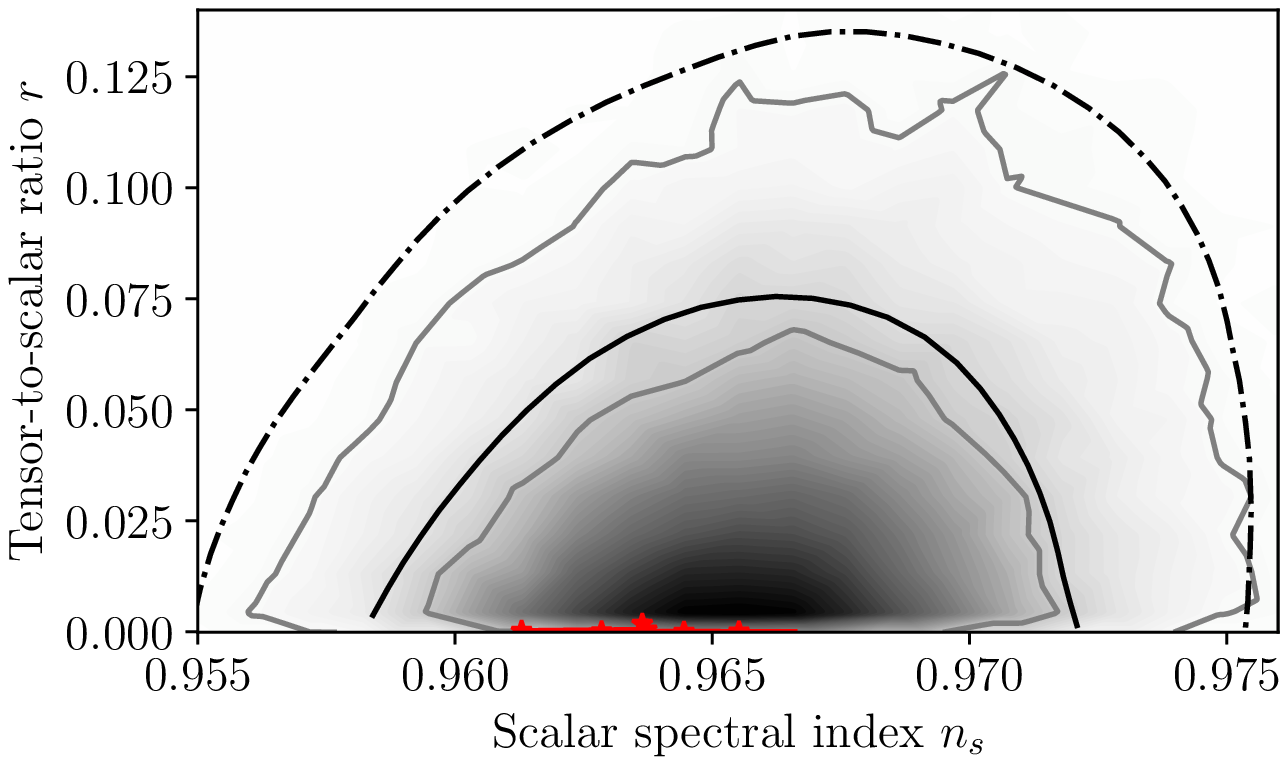}
\includegraphics[width=18pc]{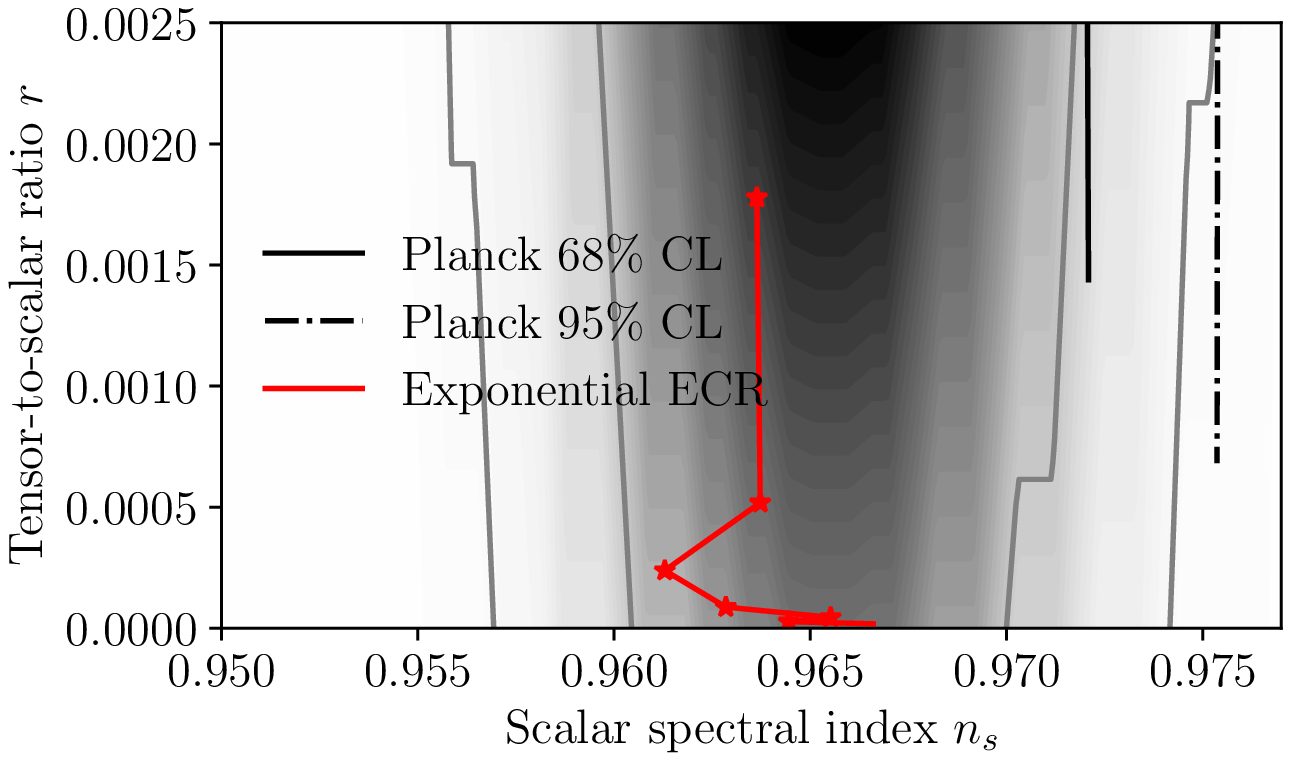}
\caption{The extended constant-roll simple exponential inflation
scenario (red lines and points) in view of the Planck 2018 data.
The bottom plot is a closeup of the upper plot.} \label{plot3}
\end{figure}

Let us now consider the linear power-law inflation scenario, which
has the following scalar potential, $V(\phi)=\frac{V_0}{\kappa^4}
\kappa \phi$, where $V_0$ again as the dimensionless potential
multiplication factor the value of is irrelevant for our analysis.
The dimensionless function $\alpha (\phi)$ of the extended
constant-roll condition (\ref{extendedconstantrollcondition1}) is
again chosen as, $\alpha (\phi)=\delta \left(\kappa \phi
\right)^m-1$. The slow-roll indices for the linear inflation model
take the simple form $\epsilon_1=\frac{1}{2} \delta ^2 \kappa ^{2
m-2} \phi ^{2 m-2}$, and $\epsilon_2=\frac{3 \kappa ^{-m} \phi
^{-m}}{\delta }-3$. As in the previous section, the values of the
scalar field at the end of inflation $\phi_f$, and at horizon
crossing $\phi_i$ are easily found to be,
$\phi_f=\frac{2^{\frac{1}{2 m-2}} \delta ^{\frac{1}{1-m}}}{\kappa
}$, $\phi_i=\left(2^{-\frac{m-2}{2 (m-1)}} \delta
^{\frac{m-2}{m-1}} \kappa ^{m-2}+\delta  (m-2) N \kappa
^{m-2}\right)^{\frac{1}{2-m}}$. Accordingly, the observational
indices take the following forms, $n_s= -2 \delta ^2 \kappa ^{2
m-2} \phi ^{2 m-2}-\frac{6 \kappa ^{-m} \phi ^{-m}}{\delta }+7$,
and $r= 8 \delta ^2 \kappa ^{2 m-2} \phi ^{2 m-2}$. Upon
evaluation of the observational indices at first horizon crossing,
we can directly confront the model at hand with the Planck 2018
observational data. The results of our analysis for the linear
power-law inflationary model are presented in Fig. \ref{plot2}.
The free parameters values are taken in this case in the ranges
$\delta=10^{-5}-10^{-17}$ and again $m=4-12$, for $N=60$
$e$-foldings. The extended constant-roll chaotic inflation model
corresponds to the red lines and points in Fig. \ref{plot2}, while
with yellow lines and points we presented the ordinary linear
power-law slow-roll inflationary model which is known to be
incompatible with the Planck data.
Thus we demonstrated that for both the chaotic and linear
power-law inflationary models, the extended constant-roll
condition renders both viable inflationary models and compatible
with the latest Planck data. In the same spirit as in the chaotic
and linear inflation model, let us present in brief the resulting
phenomenology of a  model which fails to be compatible with the
observational data in the standard slow-roll description. Let us
consider the simple exponential model, with potential,
$V(\phi)=\frac{V_0}{\kappa^4} \exp \left(\lambda \kappa
\phi\right)$ where $\lambda$ is a dimensionless parameter. In this
case we choose the dimensionless function $\alpha (\phi )$ to be,
$\alpha (\phi)=\left(\kappa \phi \right)^m-1$, and note the
absence of the parameter $\delta$ which was present in the
previous two models. The results of our analysis for the
exponential model are presented in Fig. \ref{plot3}. The free
parameters values are in the ranges $\lambda=8.1\times
10^{-3}-10^{-3}$ and $m=4-12$, for $N=60$ $e$-foldings. The
extended constant-roll exponential inflation model corresponds to
the red lines and points in Fig. \ref{plot3}, where the bottom
plot is a closeup of the upper plot. The slow-roll exponential
model yields a constant slow-roll index $\epsilon_1$ and is quite
problematic, however the extended constant-roll exponential model
is compatible with the Planck data for all the values of the free
parameters we used. As it can be seen, the model predicts
inflation deeply in the Planck $95\%$ CL likelihood curve.

\section{Symmetries and Tracking Condition Framework}

We have named the condition (\ref{extendedconstantrollcondition1})
as extended constant-roll condition, however we have not properly
justified why we used the terminology extended constant-roll
condition. The constant-roll condition is basically,
\begin{equation}\label{constantrollcondactual}
\ddot{\phi}=\beta H\dot{\phi}\, ,
\end{equation}
where $\beta$ is some constant number. Let us now see how our
terminology extended constant-roll for the condition
(\ref{extendedconstantrollcondition1}) is justified. Before we do
that, let us present an alternative definition of the extended
constant-roll condition, which can be materialized by the
following condition,
\begin{equation}\label{extendedconstantrollcondition1eq}
\ddot{\phi}=\beta (\phi) H \dot{\phi}\, ,
\end{equation}
where $\beta (\phi)$ is some smooth dimensionless function of the
scalar field $\phi$. The relation
(\ref{extendedconstantrollcondition1eq}) is a direct
generalization of the constant-roll condition
(\ref{constantrollcondactual}), for $\beta(\phi)=\mathrm{const}$.
In view of the extended constant-roll condition
(\ref{extendedconstantrollcondition1eq}), the equation of motion
for the scalar field (\ref{motion3a}) yields,
\begin{equation}\label{modifiedscalareqneq}
\dot{\phi}=-\frac{V'(\phi)}{\left( 3+\beta (\phi)\right) H}\, .
\end{equation}
Now again assuming the standard slow-roll condition for the scalar
field, namely Eq. (\ref{slowrollconditionscalarfield}) the
Friedmann equation takes the form (\ref{friedmannupdate}), hence
by using Eqs. (\ref{modifiedscalareqneq}) and
(\ref{friedmannupdate}), the slow-roll indices
(\ref{slowrollindicesdef}) take the following compact forms in
this case,
\begin{align}\label{slowrollindicesdefeq}
& \epsilon_1=\frac{9}{2\kappa^2\left(3+\beta(\phi) \right)^2}\left(\frac{V'}{V}\right)^2\\
\notag & \epsilon_2=\beta (\phi)\, .
\end{align}
Also, the $e$-foldings number for the case at hand, can easily be
found by using Eq. (\ref{modifiedscalareqneq}) that,
\begin{equation}\label{efoldingsextendedconstantrolleq}
N=\int_{\phi_i}^{\phi_f}\frac{H}{\dot{\phi}}d\phi=\kappa^2\int_{\phi_f}^{\phi_i}\frac{V}{3V'}\left(3+\beta
(\phi) \right) d\phi\, .
\end{equation}
At this point we shall show how the terminology extended
constant-roll is justified for the condition
(\ref{extendedconstantrollcondition1}). Basically the inflationary
phenomenologies related to the conditions
(\ref{extendedconstantrollcondition1}) and
(\ref{extendedconstantrollcondition1eq}) are directly related by
using the following transformation,
\begin{equation}\label{transformation}
\beta (\phi)\to -\frac{3\alpha (\phi)}{1+\alpha (\phi)}\, .
\end{equation}
It is easy to see that the two phenomenologies are identical under
the transformation (\ref{transformation}), and this is why we used
the terminology extended constant-roll for the condition
(\ref{extendedconstantrollcondition1}), since the physics of this
condition is identical to the physics of condition
(\ref{extendedconstantrollcondition1eq}).

Before closing, it is worth mentioning an interesting perspective
of the extended constant-roll scenario, related to tracking
solutions \cite{Steinhardt:1999nw}. Tracker fields or tracking
attractor solutions were initially introduced in order to solve
the coincidence problem in quintessence theories. Tracker fields
have an equation of motion with attractor solutions which rapidly
converge to a common universal attractor for a wide range of
initial conditions. Tracker solutions are appealing due to the
fact that the scalar field energy eventually overtakes the energy
density of dark matter eventually at late times. An important
condition in order to have tracking solutions is the following
\cite{Steinhardt:1999nw},
\begin{equation}\label{trackercondition}
\frac{V'}{V}=\frac{H}{\dot{\phi}}\, ,
\end{equation}
which is known as tracker condition. In a quintessence theory it
would be desirable to describe with the same scalar field both
early and late times. However as we now demonstrate, the tracker
condition eventually destroys slow-roll dynamics. On the contrary,
the tracker condition has some interesting physics to offer if it
is considered in conjunction with the extended constant-roll
condition. Let us consider first the standard slow-roll approach,
in which case the following two conditions govern the dynamics of
the scalar field,
\begin{equation}\label{slowrollstandard}
\frac{\dot{\phi}^2}{2}\ll V,\,\,\,\ddot{\phi}\ll H \dot{\phi}\, .
\end{equation}
In view of the standard slow-roll conditions
(\ref{slowrollstandard}), the equation of motion for the scalar
field (\ref{motion3a}) yields,
\begin{equation}\label{phidotslowroll}
V'\simeq -3H\dot{\phi}\, .
\end{equation}
The tracker condition (\ref{trackercondition}) can be written as,
\begin{equation}\label{reformedtracker}
\frac{V'}{V}=\frac{H\dot{\phi}}{\dot{\phi}^2}\, ,
\end{equation}
and since $V'$ and $H\dot{\phi}$ are of the same order due to Eq.
(\ref{phidotslowroll}), Eq. (\ref{reformedtracker}) implies that
$V$ and $\dot{\phi}^2$ are of the same order, which contradicts
the first slow-roll condition of Eq. (\ref{slowrollstandard}).
Thus the tracker condition (\ref{trackercondition}) is not
compatible with the slow-roll evolution, and this indicate it is
rather hard to describe with the same scalar field inflation and
quintessence tracker solutions. The same conclusion can be
obtained easily if we directly substitute Eq.
(\ref{phidotslowroll}) in Eq. (\ref{trackercondition}), to obtain,
$\frac{1}{2\kappa^2}\left(\frac{V'}{V}\right)^2\simeq
-\frac{1}{2}$, where we used that during slow-roll $3H^2\simeq
\kappa^2 V$.

Now let us demonstrate the implications of the tracker conditions
on a single scalar field theory cosmology which evolves with the
extended constant-roll condition holding true. In the extended
constant-roll evolution, the term $\dot{\phi}$ is given by Eq.
(\ref{modifiedscalareqn}), thus upon substituting this in the
tracker condition (\ref{trackercondition}), we obtain,
\begin{equation}\label{extendedtracker}
\frac{3H^2V}{\left(V'\right)^2}=-\left(1+\alpha (\phi) \right)\, ,
\end{equation}
or by using $3H^2=\kappa^2\,V$, and recalling the functional form
of the first slow-roll index $\epsilon_1$ from Eq.
(\ref{slowrollindicesdef}), the tracker condition
(\ref{trackercondition}) is transformed to a condition for the
slow-roll index $\epsilon_1$, which is,
\begin{equation}\label{trackerexspilon1}
\epsilon_1=-\frac{1+\alpha (\phi)}{2}\, .
\end{equation}
Essentially the tracker condition constrains the forms of the
potential in such a way so that the first slow-roll index
$\epsilon_1$ takes the simple form (\ref{trackerexspilon1}). The
second slow-roll index remains identical with the one appearing in
Eq. (\ref{slowrollindicesdef}), and what further changes is the
final expression for the $e$-foldings number, which actually
further constrains the functional form that $\alpha (\phi)$ can
take. Actually, in view of the tracker condition
(\ref{extendedtracker}), the $e$-foldings number expression
(\ref{efoldingsextendedconstantroll}) takes the following simple
form,
\begin{equation}\label{efoldingsnumbertracker}
N=\kappa\int_{\phi_i}^{\phi_f}\frac{d \phi}{\sqrt{-1-\alpha
(\phi)}}\, ,
\end{equation}
with the constraint that $\alpha (\phi)+1<0$. In principle the
tracker condition does not contradict scalar field cosmologies
which satisfy the extended constant-roll condition. Thus basically
the tracker condition in conjunction with the extended
constant-roll condition, introduce a new class of inflationary
phenomenology, which can combine both the tracking condition and
the extended constant-roll conditions. This class of theory could
be of importance if the unification of inflation with late-time
acceleration is the aim of a theory. We have not further analyzed
this newly introduced class of models, however, we aim to further
develop these theories in a future work. Finally, let us briefly
comment that even the constant-roll condition leads to problems if
it is considered in conjunction with the tracker condition
(\ref{trackercondition}). Indeed, for $\ddot{\phi}=\beta H
\dot{\phi}$ ($\beta$ recall is constant), the tracker condition in
conjunction with the equation of motion of the scalar field yields
$\epsilon_1=-\frac{3}{2}\frac{1}{3+\beta}$, which is not
incompatible in general with the constant-roll condition. However
this case is obviously problematic because both $\epsilon_1$ and
$\epsilon_2$ are basically constants. Thus the inflationary
phenomenology of the resulting theory is severely restricted, and
no new insights are expected. On the contrary, the tracker
condition in conjunction with the extended constant-roll condition
may offer the possibility of having late-time evolution with a de
Sitter equation of state parameter, and also describe inflation
with the same theory. This issue will be addressed in a future
work. Finally, let us note that in the context of the extended
constant-roll framework considered in conjunction with the tracker
condition, the scalar field potential and the scalar function
$\alpha (\phi)$ are no longer freely chosen since these are
connected by Eq. (\ref{slowrollindicesdef}), so basically the
potential $V(\phi)$ and the function $\alpha (\phi)$ must satisfy,
\begin{equation}\label{basicequation}
\kappa^2\left(\frac{V}{V'}\right)^2=-(1+\alpha (\phi))\, .
\end{equation}

\section{Conclusions}

In this paper we introduced the theoretical framework of extended
constant-roll dynamics for minimally coupled single scalar field
cosmologies. The extended constant-roll condition constrains
basically the scalar field evolution which was required to satisfy
the condition $\ddot{\phi}=\alpha (\phi) V'(\phi)$. Accordingly we
derived the corresponding field equations, and we calculated the
inflationary indices and the corresponding observational indices
of inflation, focusing on the spectral index of the primordial
scalar curvature perturbations and the tensor-to-scalar ratio.
Employing the new formalism by using a simple power-law $\alpha
(\phi)$ function, we demonstrated that three otherwise problematic
scalar potentials in the context of slow-roll inflation, become
compatible with the latest Planck observational data in the
context of the extended constant-roll framework. Particularly, we
studied the chaotic inflation scenarios, an exponential potential
and finally a linear power-law potential. The condition
$\ddot{\phi}=\alpha (\phi) V'(\phi)$ is equivalent to the
condition  $\ddot{\phi}=\beta (\phi) H \dot{\phi}$, which is a
direct extension of the original constant-roll condition, for
non-constant functions $\beta (\phi)$. Finally, we imposed an
extra condition to be satisfied by the extended constant-roll
scalar fields, namely the tracker condition. Originally the
tracker condition was engineered for the quintessence scenario in
order to explain the coincidence problem. As we showed, the
extended constant-roll scenario is compatible with the tracker
condition, and a new inflationary phenomenological framework
arises with this new constraint, in contrast to the slow-roll and
constant-roll cases, which lead to inconsistencies if the tracker
condition is imposed. We briefly discussed how the inflationary
phenomenology can be extracted with this new tracker condition
constraints, the main feature of which is that the function
$\alpha (\phi)$ and the potential $V(\phi)$ are directly
functionally related. In a future work a compelling task would be
to investigate whether the same scalar field can simultaneously
describe inflation and the dark energy era, and we hope to address
this issue in a future work.


\begin{thebibliography}{99}




\bibitem{Guth:1980zm}
A.~H.~Guth,
Phys.\ Rev.\ D {\bf 23} (1981) 347. doi:10.1103/PhysRevD.23.347

\bibitem{Starobinsky:1982ee}
A.~A.~Starobinsky,
Phys.\ Lett.\ {\bf 91B} (1980) 99.
doi:10.1016/0370-2693(80)90670-X

\bibitem{Linde:1983gd}
A.~D.~Linde,
Phys.\ Lett.\ {\bf 129B} (1983) 177.
doi:10.1016/0370-2693(83)90837-7


\bibitem{Albrecht:1982wi}
A.~Albrecht and P.~J.~Steinhardt,
Phys.\ Rev.\ Lett.\  {\bf 48} (1982) 1220 [Adv.\ Ser.\ Astrophys.\
Cosmol.\  {\bf 3} (1987) 158]. doi:10.1103/PhysRevLett.48.1220






\bibitem{Nojiri:2017ncd}
S.~Nojiri, S.~D.~Odintsov and V.~K.~Oikonomou,
Phys.\ Rept.\ {\bf 692} (2017) 1 doi:10.1016/j.physrep.2017.06.001
[arXiv:1705.11098 [gr-qc]].

\bibitem{Nojiri:2010wj}
S.~Nojiri and S.~D.~Odintsov,
Phys.\ Rept.\ {\bf 505} (2011) 59
doi:10.1016/j.physrep.2011.04.001 [arXiv:1011.0544 [gr-qc]].

\bibitem{Nojiri:2006ri}
S.~Nojiri and S.~D.~Odintsov,
eConf C {\bf 0602061} (2006) 06
 [Int.\ J.\ Geom.\ Meth.\ Mod.\ Phys.\ {\bf 4} (2007) 115]
doi:10.1142/S0219887807001928 [hep-th/0601213].

\bibitem{Capozziello:2011et}
S.~Capozziello and M.~De Laurentis,
Phys.\ Rept.\ {\bf 509} (2011) 167
doi:10.1016/j.physrep.2011.09.003 [arXiv:1108.6266 [gr-qc]].

\bibitem{Capozziello:2010zz}
V.~Faraoni and S.~Capozziello,
Fundam.\ Theor.\ Phys.\ {\bf 170} (2010).
doi:10.1007/978-94-007-0165-6

\bibitem{delaCruzDombriz:2012xy}
A.~de la Cruz-Dombriz and D.~Saez-Gomez,
Entropy {\bf 14} (2012) 1717 doi:10.3390/e14091717
[arXiv:1207.2663 [gr-qc]].

\bibitem{Olmo:2011uz}
G.~J.~Olmo,
Int.\ J.\ Mod.\ Phys.\ D {\bf 20} (2011) 413
doi:10.1142/S0218271811018925 [arXiv:1101.3864 [gr-qc]].






\bibitem{Akrami:2018odb}
Y.~Akrami {\it et al.} [Planck Collaboration],
arXiv:1807.06211 [astro-ph.CO].









\bibitem{Inoue:2001zt}
S.~Inoue and J.~Yokoyama,
Phys.\ Lett.\ B {\bf 524} (2002) 15
doi:10.1016/S0370-2693(01)01369-7 [hep-ph/0104083].

\bibitem{Tsamis:2003px}
N.~C.~Tsamis and R.~P.~Woodard,
Phys.\ Rev.\ D {\bf 69} (2004) 084005
doi:10.1103/PhysRevD.69.084005 [astro-ph/0307463].

\bibitem{Kinney:2005vj}
W.~H.~Kinney,
Phys.\ Rev.\ D {\bf 72} (2005) 023515
doi:10.1103/PhysRevD.72.023515 [gr-qc/0503017].

\bibitem{Tzirakis:2007bf}
K.~Tzirakis and W.~H.~Kinney,
Phys.\ Rev.\ D {\bf 75} (2007) 123510
doi:10.1103/PhysRevD.75.123510 [astro-ph/0701432].

\bibitem{Namjoo:2012aa}
M.~H.~Namjoo, H.~Firouzjahi and M.~Sasaki,
Europhys.\ Lett.\  {\bf 101} (2013) 39001
doi:10.1209/0295-5075/101/39001 [arXiv:1210.3692 [astro-ph.CO]].

\bibitem{Martin:2012pe}
J.~Martin, H.~Motohashi and T.~Suyama,
Phys.\ Rev.\ D {\bf 87} (2013) no.2,  023514
doi:10.1103/PhysRevD.87.023514 [arXiv:1211.0083 [astro-ph.CO]].

\bibitem{Motohashi:2014ppa}
H.~Motohashi, A.~A.~Starobinsky and J.~Yokoyama,
JCAP {\bf 1509} (2015) no.09,  018
doi:10.1088/1475-7516/2015/09/018 [arXiv:1411.5021 [astro-ph.CO]].

\bibitem{Cai:2016ngx}
Y.~F.~Cai, J.~O.~Gong, D.~G.~Wang and Z.~Wang,
JCAP {\bf 1610} (2016) no.10,  017
doi:10.1088/1475-7516/2016/10/017 [arXiv:1607.07872
[astro-ph.CO]].

\bibitem{Motohashi:2017aob}
H.~Motohashi and A.~A.~Starobinsky,
arXiv:1702.05847 [astro-ph.CO].

\bibitem{Hirano:2016gmv}
S.~Hirano, T.~Kobayashi and S.~Yokoyama,
Phys.\ Rev.\ D {\bf 94} (2016) no.10,  103515
doi:10.1103/PhysRevD.94.103515 [arXiv:1604.00141 [astro-ph.CO]].

\bibitem{Anguelova:2015dgt}
L.~Anguelova,
Nucl.\ Phys.\ B {\bf 911} (2016) 480
doi:10.1016/j.nuclphysb.2016.08.020 [arXiv:1512.08556 [hep-th]].

\bibitem{Cook:2015hma}
J.~L.~Cook and L.~M.~Krauss,
JCAP {\bf 1603} (2016) no.03,  028
doi:10.1088/1475-7516/2016/03/028 [arXiv:1508.03647
[astro-ph.CO]].

\bibitem{Kumar:2015mfa}
K.~S.~Kumar, J.~Marto, P.~Vargas Moniz and S.~Das,
JCAP {\bf 1604} (2016) no.04,  005
doi:10.1088/1475-7516/2016/04/005 [arXiv:1506.05366 [gr-qc]].

\bibitem{Odintsov:2017yud}
S.~D.~Odintsov and V.~K.~Oikonomou,
arXiv:1703.02853 [gr-qc].

\bibitem{Odintsov:2017qpp}
S.~D.~Odintsov and V.~K.~Oikonomou,
arXiv:1704.02931 [gr-qc].

\bibitem{Lin:2015fqa}
J.~Lin, Q.~Gao and Y.~Gong,
Mon.\ Not.\ Roy.\ Astron.\ Soc.\  {\bf 459} (2016) no.4,  4029
doi:10.1093/mnras/stw915 [arXiv:1508.07145 [gr-qc]].

\bibitem{Gao:2017uja}
Q.~Gao and Y.~Gong,
arXiv:1703.02220 [gr-qc].



\bibitem{Nojiri:2017qvx}
  S.~Nojiri, S.~D.~Odintsov and V.~K.~Oikonomou,
  Class.\ Quant.\ Grav.\  {\bf 34} (2017) no.24,  245012
  doi:10.1088/1361-6382/aa92a4
  [arXiv:1704.05945 [gr-qc]].





\bibitem{Oikonomou:2017bjx}
  V.~K.~Oikonomou,
  Mod.\ Phys.\ Lett.\ A {\bf 32} (2017) no.33,  1750172
  doi:10.1142/S0217732317501723
  [arXiv:1706.00507 [gr-qc]].




\bibitem{Odintsov:2017hbk}
  S.~D.~Odintsov, V.~K.~Oikonomou and L.~Sebastiani,
  Nucl.\ Phys.\ B {\bf 923} (2017) 608
  doi:10.1016/j.nuclphysb.2017.08.018
  [arXiv:1708.08346 [gr-qc]].



\bibitem{Oikonomou:2017xik}
  V.~K.~Oikonomou,
  Int.\ J.\ Mod.\ Phys.\ D {\bf 27} (2017) no.02,  1850009
  doi:10.1142/S0218271818500098
  [arXiv:1709.02986 [gr-qc]].



\bibitem{Cicciarella:2017nls}
  F.~Cicciarella, J.~Mabillard and M.~Pieroni,
  JCAP {\bf 1801} (2018) no.01,  024
  doi:10.1088/1475-7516/2018/01/024
  [arXiv:1709.03527 [astro-ph.CO]].




\bibitem{Awad:2017ign}
  A.~Awad, W.~El Hanafy, G.~G.~L.~Nashed, S.~D.~Odintsov and V.~K.~Oikonomou,
  JCAP {\bf 1807} (2018) no.07,  026
  doi:10.1088/1475-7516/2018/07/026
  [arXiv:1710.00682 [gr-qc]].





\bibitem{Anguelova:2017djf}
  L.~Anguelova, P.~Suranyi and L.~C.~R.~Wijewardhana,
  JCAP {\bf 1802} (2018) no.02,  004
  doi:10.1088/1475-7516/2018/02/004
  [arXiv:1710.06989 [hep-th]].




\bibitem{Ito:2017bnn}
  A.~Ito and J.~Soda,
  Eur.\ Phys.\ J.\ C {\bf 78} (2018) no.1,  55
  doi:10.1140/epjc/s10052-018-5534-5
  [arXiv:1710.09701 [hep-th]].



\bibitem{Karam:2017rpw}
  A.~Karam, L.~Marzola, T.~Pappas, A.~Racioppi and K.~Tamvakis,
  JCAP {\bf 1805} (2018) no.05,  011
  doi:10.1088/1475-7516/2018/05/011
  [arXiv:1711.09861 [astro-ph.CO]].

\bibitem{Yi:2017mxs}
  Z.~Yi and Y.~Gong,
  JCAP {\bf 1803} (2018) no.03,  052
  doi:10.1088/1475-7516/2018/03/052
  [arXiv:1712.07478 [gr-qc]].





\bibitem{Mohammadi:2018oku}
  A.~Mohammadi, K.~Saaidi and T.~Golanbari,
  Phys.\ Rev.\ D {\bf 97} (2018) no.8,  083006
  doi:10.1103/PhysRevD.97.083006
  [arXiv:1801.03487 [hep-ph]].




\bibitem{Gao:2018tdb}
  Q.~Gao, Y.~Gong and Q.~Fei,
  JCAP {\bf 1805} (2018) no.05,  005
  doi:10.1088/1475-7516/2018/05/005
  [arXiv:1801.09208 [gr-qc]].





\bibitem{Mohammadi:2018wfk}
  A.~Mohammadi and K.~Saaidi,
  arXiv:1803.01715 [astro-ph.CO].




\bibitem{Morse:2018kda}
  M.~J.~P.~Morse and W.~H.~Kinney,
  Phys.\ Rev.\ D {\bf 97} (2018) no.12,  123519
  doi:10.1103/PhysRevD.97.123519
  [arXiv:1804.01927 [astro-ph.CO]].



\bibitem{Cruces:2018cvq}
  D.~Cruces, C.~Germani and T.~Prokopec,
  JCAP {\bf 1903} (2019) no.03,  048
  doi:10.1088/1475-7516/2019/03/048
  [arXiv:1807.09057 [gr-qc]].


\bibitem{GalvezGhersi:2018haa}
  J.~T.~Galvez Ghersi, A.~Zucca and A.~V.~Frolov,
  JCAP {\bf 1905} (2019) no.05,  030
  doi:10.1088/1475-7516/2019/05/030
  [arXiv:1808.01325 [astro-ph.CO]].



\bibitem{Boisseau:2018rgy}
  B.~Boisseau and H.~Giacomini,
  arXiv:1809.09169 [gr-qc].



\bibitem{Gao:2019sbz}
  Q.~Gao, Y.~Gong and Z.~Yi,
  arXiv:1901.04646 [gr-qc].


\bibitem{Lin:2019fcz}
  W.~C.~Lin, M.~J.~P.~Morse and W.~H.~Kinney,
  arXiv:1904.06289 [astro-ph.CO].








\bibitem{Hwang:2005hb}
  J.~c.~Hwang and H.~Noh,
  Phys.\ Rev.\ D {\bf 71} (2005) 063536
  doi:10.1103/PhysRevD.71.063536
  [gr-qc/0412126].





\bibitem{DeFelice:2011zh}
  A.~De Felice and S.~Tsujikawa,
  JCAP {\bf 1104} (2011) 029
  doi:10.1088/1475-7516/2011/04/029
  [arXiv:1103.1172 [astro-ph.CO]].



\bibitem{Steinhardt:1999nw}
P.~J.~Steinhardt, L.~M.~Wang and I.~Zlatev,
Phys. Rev. D \textbf{59} (1999), 123504
doi:10.1103/PhysRevD.59.123504 [arXiv:astro-ph/9812313
[astro-ph]].






\end{thebibliography}
\end{document}